\documentclass[aps,prd,twocolumn,groupedaddress,showpacs,nofootinbib]{revtex4}
\usepackage{amsmath,amssymb}\begin{document}
\title{Classification of cosmological milestones}
\author{L. Fern\'andez-Jambrina}
\email[]{leonardo.fernandez@upm.es}
\homepage[]{http://debin.etsin.upm.es/ilfj.htm}
\affiliation{Matem\'atica Aplicada, E.T.S.I. Navales, Universidad
Polit\'ecnica de Madrid,\\
Arco de la Victoria s/n, \\ E-28040 Madrid, Spain}%
\author{Ruth Lazkoz}\email[]{ruth.lazkoz@ehu.es} \affiliation{F\'\i
sica Te\'orica, Facultad de Ciencia y Tecnolog\'\i a, Universidad del
Pa\'\i s Vasco,\\ Apdo.  644, E-48080 Bilbao, Spain}
      
\begin{abstract}In this paper causal geodesic completeness of FLRW
cosmological models is analysed in terms of generalised power
expansions of the scale factor in coordinate time.  The strength of
the found singularities is discussed following the usual definitions
due to Tipler and Kr\'olak.  It is shown that while classical
cosmological models are both timelike and lightlike geodesically
incomplete, certain observationally alllowed models which have been
proposed recently are lightlike geodesically complete.\end{abstract}

\maketitle

\section{Introduction}

In the last few years the evidence that the Universe contains a
large proportion
of some not ordinary stuff which makes it expand acceleratedly is
getting stronger grounds (see \cite{de} for a recent review). 
Due to such considerations, quite a few names have been added to the
original list of contents of the Universe along its different epochs.
Among those, quintessence and phantom energy are the most popular,
and, particularly the latter, has triggered a feverish activity in many
directions, one of them being the investigation of the unusual
geometric properties of the cosmological models they lead to. Phantom
universes
have rather awkward singularities \cite{caldwell}, which go into the
basket of ``cosmological milestones'' \cite{visser} along with other
geometric curiosities such as bounces or extremality events
appearing in other sorts of universes. 

Observationally, phantom universes seem to be preferred over
geometry-wise dull LCDM cosmologies \cite{obs}, thus making
legitimate the sort of questions addressed here  about the
fate of phantom  universes. We are going  to perform an innovative
analysis of those models in conjunction with all the other FLRW
models in the literature, which will bring some surprises to build on
the atypicality of phantom cosmologies. In addition, our analysis
opens new paths for the exploration of other sort of milestones, such
as sudden singularities which have received considerable attention
recently. 

We have borrowed the denomination ``cosmological milestones'' from
\cite{visser}, where FLRW cosmological models were analysed in terms
of a generalised power expansion of the scale factor.  The appearance
of polynomial scalar curvature singularities and derivative curvature
singularities, together with the satisfaction of energy conditions
were shown to depend most generally on just the first three terms of
the expansion.

Clearly, there are many interesting geometrical features which are
elusive to studies of that sort.  Since the usual definitions of
singularities are related not only to the properties of the curvature
tensor but also to the existence of causal geodesics that cannot be
extended to arbitrary values of their proper time (geodesic
incompleteness) \cite{HE} or even of general causal curves with the
same property (b-incompleteness) \cite{ehresman, schmidt}, it is of
interest to analyse singularities in general FLRW cosmologies within
this framework, as it was done in \cite{flrw} for sudden singularities
\cite{barrow}.  It is relevant to do so because causal geodesics
describe the trajectories of observers subject just to gravitational
forces.  Note that curvature is a static concept, in the sense that it
only reflects what happens at each event, whereas features derived
from tracking the observer along its trajectory are more dynamical,
and somewhat more enlightening.  Thus, our study covers key issues
that were overlooked in recent related classifications \cite{visser,
classodi}

We begin therefore in section \ref{eqs} by arranging geodesic
equations for FLRW cosmological models in a suitable fashion for
integration in terms of a generalised power expansion in coordinate
time.  We proceed then in section \ref{light} to analyse the behavior
of lightlike geodesics in these models, which sets the foundations
allowing to check whether the singularities that are found are strong
or not according to the usual definitions reviewed and refined in
section \ref{stronglight}.  Timelike geodesics are integrated in
section \ref{time} and the strength of their singularities is dealt
with in section \ref{strongtime}.  The paper ends with a discussion of
the results in section \ref{discuss}.  Special remarks are done
throughout the paper regarding observationally allowed/favoured
phantom cosmologies, because there is a peculiar class of such
cosmologies which persist to stand out of the crowd of all phantom
models, as long as their geometrical properties are concerned.

\section{Geodesics in FLRW cosmological models\label{eqs}}

We  consider spacetimes endowed with a FLRW metric of the form
\begin{eqnarray}
&&ds^2=-dt^2+a^2(t)\left\{f^2(r)dr^2+ r^2\left(d\theta^2+\sin^2\theta
d\phi^2\right)\right\}\nonumber\\
&&f^2(r)=\frac{1}{1-kr^2},\quad k=0,\pm1,\label{metric}\end{eqnarray}

As in \cite{visser}, we assume that at a coordinate time $t_{0}$, a
singular event or cosmological milestone comes up in such spacetime.
To allow our results to be most general, we just require the scale factor
$a(t)$ to have a generalised Puiseux expansion around the event at
$t_{0}$,
\begin{equation}\label{puiseux}
    a(t)=c_{0}|t-t_{0}|^{\eta_{0}}+c_{1}|t-t_{1}|^{\eta_{1}}+\cdots\;,
\end{equation}
where the exponents $\eta_{i}$ are real and ordered,
\[\eta_{0}<\eta_{1}<\cdots\]

This framework covers every proposal of FLRW cosmological model in the
literature.

At first order, a model admitting such expansion behaves as a
power-law model of exponent $\eta_{0}$, which in the case of a
flat universe, $k=0$, would correspond to a linear equation of state
for the cosmological fluid, \[p=w\rho\;,\quad w=-1+2/3 \eta_{0}\;.\]

In order to have a positive expansion factor, we require $c_{0}$ to be
positive.

Depending on the value of $\eta_{0}$, several types of cosmological
milestones arise \cite{visser}:

\begin{itemize}
    \item  $\eta_{0}>0$: the scale factor vanishes at $t_{0}$ and
    generically we have a Big Bang or Big Crunch singularity.

    \item  $\eta_{0}=0$: the scale factor is finite at $t_{0}$.
    If $a(t)$ is analytical, the event at $t_{0}$ is regular.
    Otherwise a weak or sudden singularity appears \cite{barrow}.

    \item  $\eta_{0}<0$: the scale factor diverges at $t_{0}$ and a
    Big Rip singularity appears.
\end{itemize}

Since the singular event at $t_{0}$ is approached from just one side
(the past for a Big Bang, the future for a Big Crunch singularity),
there is usually no need to consider absolute values in the expansion
(\ref{puiseux}), except in the case, for instance, of sudden or weak
singularities \cite{barrow}, which have been seen to be traversable
\cite{flrw} for geodesic observers.

In order to avoid dealing with signs, we  consider singularities 
in the past, $t>t_{0}$. Of course the analysis is valid 
also for singularities in the future, since the equations are
time-symmetric, and occasionally we will comment what would happen 
if the singularity lies in the past.

We  consider causal geodesics parametrised by their proper time
$\tau$, $\left(t(\tau),r(\tau),\theta(\tau), \phi(\tau)\right)$.  This
means that the velocity $u$ of the parametrization $(\dot t, \dot r,
\dot\theta, \dot\phi)$ satisfies
\begin{equation}\label{delta}
\delta=-g_{ij}\dot x^{i}\dot x^{j}\;, \quad x^i,
x^j=t,r,\theta,\phi\;,
\end{equation}
where $\delta$ takes the zero value for lightlike geodesics and the
value one for timelike geodesics.  It takes the value minus one for
spacelike geodesics, but since we need just causal curves for our
analysis, we will discard the $\delta=-1$ case.  The dot stands for
derivation with respect to proper time.

 Condition  (\ref{delta}) defines proper time up to a change of
scale and a translation, $\tilde\tau =A\tau + B$, and therefore the
parametrization is also called affine parametrization.

Geodesic equations are quasilinear in the acceleration $(\ddot t,
\ddot r, \ddot \theta, \ddot \phi)$ and depend on the metric
components $g_{ij}$ through the Christoffel symbols,
\begin{equation}
\ddot x^{i}+\Gamma^i_{jk}\dot x^j\dot x^k=0\;,\end{equation}
\begin{equation}
\Gamma^{i}_{jk}=\frac{1}{2}g^{il}\left\{g_{lj,k}+g_{lk,j}-g_{jk,l}\right\}\;.\end{equation}

For a FLRW cosmology they may be written as
\begin{subequations}
    \begin{eqnarray}
\ddot t&=&-\frac{aa'}{1-kr^2}\dot
r^2-aa'r^2\left(\dot\theta^2+\sin^2\theta\dot\phi^2\right)\\
\ddot r&=&-2\frac{a'}{a}\dot t\dot r-\frac{kr}{1-kr^2}\dot
r^2\nonumber\\&+&
(1-kr^2)r(\dot\theta^2+\sin^2\theta\dot\phi^2)\\
\ddot\theta&=&-2\frac{a'}{a}\dot t\dot\theta-\frac{2}{r}\dot
r\dot\theta
+\sin\theta\cos\theta\dot\phi^2
\\
\ddot\phi&=&-2\frac{a'}{a}\dot t\dot\phi-\frac{2}{r}\dot r\dot\phi-
2\cot\theta\dot\theta\dot\phi\;,
\end{eqnarray}
	\end{subequations}
where the comma stands for derivation with respect to coordinate time
$t$. 

Taking into account that orbits of geodesics in spherically
symmetric spacetimes remain in equatorial hypersurfaces, they can be
fit in a
hypersurface $\theta=\pi/2$ by choosing coordinates accordingly. This 
allows a simplication of the system of equations,
\begin{subequations}\label{geoeqs}
    \begin{eqnarray}
\ddot t&=&-\frac{aa'}{1-kr^2}\dot
r^2-aa'r^2\dot\phi^2\\
\ddot r&=&-2\frac{a'}{a}\dot t\dot r-\frac{kr}{1-kr^2}\dot
r^2+(1-kr^2)r\dot\phi^2\\
\ddot\phi&=&-2\frac{a'}{a}\dot t\dot\phi-\frac{2}{r}\dot r\dot\phi\;.
\end{eqnarray}
	\end{subequations}
	
Finally, due to the existence of isometries, the 
following conserved quantities of geodesic motion exist:
\begin{subequations}\begin{eqnarray}
P_{1}&=&a(t)\left\{f(r)\cos \phi \,\dot r-\frac {
r}{f(r)}\sin \phi
\,\dot \phi\right\}\\
P_{2}&=&a(t) \left\{f(r) \sin \phi \,\dot r+\frac {
r}{f(r)}\cos \phi\,\dot \phi\right\}\\
L&=&a(t)r^2\,\dot\phi\;.
\end{eqnarray}
\end{subequations}
They allow us to reduce (\ref{geoeqs})  to a simple set of
first order differential equations:
\begin{subequations}\label{geod}
\begin{eqnarray}\dot t^2&=&\delta +\label{geod1}
\frac{P^2+k L^2}{a^2(t)}\;,\\\dot r&=&\frac {P_{1}\,\cos \phi +P_{2}\,
\sin \phi}{a^2(t)f(r)}\;,\label{geod2}\\\dot\phi&=&\frac
{L}{a^2(t)r^2}\;.
\end{eqnarray}
\end{subequations}

These constants of geodesic motion are related to angular
momentum and linear momentum,
\begin{equation}
P^2=P_{1}^2+P_{2}^2\;,\end{equation} and even allow us to obtain the
equation of the orbits, namely,
\begin{equation}
    f(r)\,r=\frac{L}{P_{2}\cos\phi-P_{1}\sin\phi}\;,
    \end{equation}
which is just the equation of a straight line in polar coordinates in 
the flat case $k=0$:
\[P_{2}x-P_{1}y= L\;,\]as it was to be expected.

It may be seen that the distinction between different types of universes
(flat, open, closed) appears only through the constant $k$ in
(\ref{geod1}) and in the function $f(r)$ in (\ref{geod2}).  This
function may be factored away in that equation just by taking \[\dot
R=\frac {P_{1}\,\cos \phi +P_{2}\, \sin \phi}{a^2(t)}\;, \quad
R=\left\{
\begin{array}{ll}
    \mathrm{arcsinh}\, r & k=-1  \\
    r & k=0  \\
    \arcsin r & k=1
\end{array}
\right..\]

Therefore, the relevant information for geodesics is encoded in the
scale factor $a(t)$. We  make use of this fact on analysing the
behaviour of causal geodesics.

From (\ref{geod}) we learn that just the equation for $t$ needs to be
solved, since the equations for the other coordinates are reduced to
quadratures once the solution of (\ref{geod1}) is known.

We can forget the equation for $\phi$ along a geodesic since, due to
homogeneity and isotropy of the FLRW universe, the origin may be
located at any point on the geodesic and hence this appears as a
straight line with zero angular velocity:
\begin{subequations}
\begin{eqnarray}\label{geods}
    \dot t&=&\sqrt{\delta +\frac{P^2}{a^2(t)}}\;,\label{geods1}\\\dot
    r&=&\pm\frac {P}{a^2(t)f(r)}
    \;.\label{geods2}\end{eqnarray}
    \end{subequations}

Possibly a quicker way to reach this result is considering beforehand 
that geodesics are straight lines due to the homogeneity and isotropy 
of the spacetime and that $\partial_{R}=\partial_{r}/f(r)$ is a
generator of an isometry
along one of these lines. Hence, 
\[\pm P=u\cdot\frac{\partial_{r}}{f(r)}=a^2(t)f(r)\dot r\;,\]is a
conserved quantity
of geodesic motion. The equation for $\dot t$ is derived from the
normalization condition (\ref{delta}),
\[\delta =\dot t^2-a^2(t)f^2(r) \dot r^2\;.\]

We  consider future-pointing geodesics and therefore we take
$\dot t>0$.

Now we may begin to draw information about causal geodesics from their
equations.  Following \cite{HE} we take causal geodesic completeness
as a minimum
condition for a spacetime to be considered singularity-free.
Therefore, we  analyse the cases where causal geodesics are
incomplete, that is, where they cannot be extended to
arbitrarily large
values of their proper time $\tau$. 

However, we must bear in mind that, since we have no guarantee that
the coordinate chart that allows us to write the metric in the form
(\ref{metric}) covers the whole universe, some conclusions about
incompleteness may not be correct if the spacetime is extendible to a
larger one.  That is, a geodesic may leave the portion of spacetime
depicted by our coordinates in finite proper time, but not the
universe itself.  Therefore, some of the singularities we may
encounter may
not be real, since the universe can be extended.

This is the case of Milne universe, which is the case of
(\ref{metric}) for $k=-1$, $a(t)=t$. This universe can be reduced to a
portion of empty Minkowski spacetime by the coordinate transformation
\[T=t\sqrt{1+r^2}\;,\quad R=rt\;,\]
which covers just the region inside the null cone $T^2=R^2$. It is,
therefore, a geodesically complete and singularity-free spacetime,
but 
it  appears singular in the Milne form, since it can be extended to
the whole Minkowski spacetime.

\section{Lightlike geodesics\label{light}}

The lightlike case is fairly simple and can be explicitly integrated
for the time coordinate,
\[ a(t)\dot t=P \Rightarrow \int_{t_{0}}^t
a(t')\,dt'=P(\tau-\tau_{0})\;.\]

Close to $t_{0}$, the leading term in the power expansion of equation 
(\ref{geods1}) is the one with lowest
exponent, $\eta_{0}$.  In many cases, in order to analyse the singular
behavior of geodesics near $t_{0}$, we just require the
first term of 
the power expansion,
\[a(t)\simeq c_{0}|t-t_{0}|^{\eta_{0}}\;,\]
which provides the time coordinate at first order in terms of proper
time $\tau$, after integrating (\ref{geods1}),
\[\int_{t_{0}}^tc_{0}|t'-t_{0}|^{\eta_{0}}\,dt'\simeq P(
\tau-\tau_{0})\;,\]
\begin{equation}\label{tlight}
t\simeq
t_{0}+\left\{\frac{(1+\eta_{0}) P}{c_{0}}\right\}^{1/(1+\eta_{0})}
(\tau-\tau_{0})^{1/(1+\eta_{0})}\;,\end{equation}
for $\eta_{0}\neq -1$. If $\eta_{0}=-1$, the leading term is
exponential,
\[t\simeq t_{0}+Ce^{P\tau/c_{0}}\;.\]  

Other cases which  require a different treatment, involving more
terms of the power expansion, are those with $\eta_0=(1-n)/n$ with
$n$ a positive natural number.  We elaborate further on those
cases below.

From either the expression of $t$ in (\ref{tlight}) or $\dot t$ and
its derivatives  one gets 
\begin{subequations}\begin{eqnarray}
    \dot t&\simeq
    &\frac{P}{c_{0}}|t-t_{0}|^{-\eta_{0}}\;,\\
    \ddot t&\simeq&-
    \frac{\eta_{0}P}{c_{0}}|t-t_{0}|^{-\eta_{0}-1}\dot
t\nonumber\\&\simeq&
    -\frac{\eta_{0}P^2}{c_{0}^2}|t-t_{0}|^{-2\eta_{0}-1}\\
    t^{n)}&\simeq&\lambda_{n}|t-t_{0}|^{1-n-n\eta_{0}}\;,\label{ltn}
  \end{eqnarray}\end{subequations}
  with
  \begin{equation}  
    \lambda_{n}=(-1)^{n+1}\eta_{0}\cdots\big((n-1)\eta_{0}+n-2\big)
{\left(
\frac{P}{c_{0}}\right)}^n\label{lambda}. 
\end{equation}

Several  possibilities arise, since (\ref{ltn}) implies that if
\begin{equation} \eta_{0}\le \frac{1-n}{n}\;,\quad
    \frac{5n-3}{3(1-n)}\le w\le -1\;,\end{equation}
there is no blow up in any of the derivatives of order lower or
equal than $n$.
The latter condition is not too stringent, that is, such cases appear often, so it is of utmost relevance to get further
insight into the details of the different subcases one can
distinguish: 
\begin{itemize}
    \item  $\eta_{0}>0$: This case includes all classical matter
    contents (for flat universes, scalar field $\eta_{0}=1/3$,
radiation $\eta_{0}=1/2$,
    dust    $\eta_{0}=2/3$\ldots, with $w>-1$). Since the exponent
$1/(1+\eta_{0})$ is
    lower than one, $t$ is not differentiable at $t_{0}$ and the
    derivative $\dot t$ blows up. 

    \item $\eta_{0}\in (-1/2,0)$: It corresponds to $w<-7/3$ for flat
    power-law models.  In this case $\dot t$ does not blow up at
    $t_{0}$, but $\ddot t$ does.
    
    \item $\eta_{0}\in (-2/3,-1/2)$: It corresponds to $w\in(-7/3,-2)$
    for flat power-law models.  In this case $\ddot t$ does not blow
up at
    $t_{0}$, but $\dddot t$ does.
    
    \item $\eta_{0}\in \big(\frac{1-n}{n},\frac{2-n}{n-1}\big)$: It
    corresponds to
$w\in\big(\frac{5n-8}{3(2-n)},\frac{5n-3}{3(1-n)}\big)$ for
    flat power-law models.  The derivative $t^{n-1)}$ does not blow
up at
    $t_{0}$ but $t^{n)}$ does.

    \item $\eta_{0}<-1$: It corresponds to $w\in(-5/3,-1)$ for
    power-law models.  According to (\ref{tlight}), the time
    coordinate along the lightlike geodesic is dominated by a negative
    power of proper time $\tau$.  This means that these geodesics
    never reach $t_{0}$, since $t-t_{0}$ only vanishes when $\tau$
    tends to infinity.  Typically, best fit phantom models have a
    value of $w$ within this range, so it seem very likely that if the
    universe is phantom its geodesics are going to have this peculiar
    behavior.  
\end{itemize}

The limit cases where $\eta_{0}$ is of the form $(1-n)/n$
(0,-1/2,-2/3, -3/4,\ldots,-1) fall out of this classification since
the derivative $t^{n+1)}$ vanishes, as it follows from (\ref{lambda}).
If every exponent $\eta_{i}$ in the generalised power expansion of
$a(t)$ were of this form, none of the derivatives of the time
coordinate along lightlike geodesics would blow up.  These are
extremely fine-tuned cases, so we will not consider them any further,
and we will then turn back to the cases $\eta_{i} =(1-n)/n$ for $i=0$
only.

 For this analysis  we  have to resort to the next
term of the power expansion, $c_{1}(t-t_{0})^{\eta_{1}}$, 
but if the term does not provide sufficient information, one would
have to keep adding terms till a satisfactory expression is
obtained.

Up to second order, \[\dot t=\frac{P}{a(t)} \simeq
\frac{P}{c_{0}}|t-t_{0}|^{-\eta_{0}}- \frac{P
c_{1}}{c_{0}^2}|t-t_{0}|^{\eta_{1}-2\eta_{0}}+\cdots\] we see that,
after the contribution of the first term to the derivative $t^{n+1)}$
vanishes, \[ t^{n+1)}\sim |t-t_{0}|^{\eta_{1}-(n+2)\eta_{0}-n}\;,\]
the leading term for lightlike geodesic behavior in these cases is the
one with $\eta_{1}$.

Let us take a look to the new cases arising:
\begin{itemize}
    \item $\eta_{0}=0$.  In this case the scale factor is neither zero
    nor it tends to infinity at $t_{0}$ and, in principle, equations
    (\ref{geods}) are regular at $t_{0}$ as it was pointed out in
    \cite{flrw}.  This is the case of the models proposed by Barrow
    \cite{barrow}.  How regular/singular geodesics are in these
    spacetimes depends on the next exponent, $\eta_{1}>0$, in the
    expansion of the scale factor.  For any such value $\dot t$ does
    not blow up at $t_{0}$, but some remarks are in order.
    
If $\eta_{1}\in \big(n-2,n-1\big)$, $n>1$, the derivative $t^{n-1)}$
does not blow up at $t_{0}$ but $t^{n)}$ does.  The phenomenon called
sudden singularity \cite{barrow}, which has aroused much
interest \cite{barrow,chilaz,lake,barrowfollow} \footnote{In
\cite{barrow} and some of the references inspired by that work
\cite{barrowfollow,barrowdab}, the treatment is purely
phenomenological and the focus is on a quest for ad hoc
parametrizations of $a(t)$ leading to such singular behavior to arise
(see \cite{chilaz} for an approximate reconstruction of the equation
of state or \cite{barrowdab} for statefinder parameters).
Interestingly, sudden singularities can appear in more solid contexts,
for instance in some braneworld models \cite{brane}.}, corresponds to
the models with $\eta_{1}>1$, which have therefore nonsingular $\ddot
t$ at $t_{0}$, and were shown to have well-defined geodesics around
$t_{0}$ in \cite{flrw}.
    
    If $\eta_{1}$ is a natural number, the same reasoning is applied
    to the first exponent $\eta_{N}$ which is not natural.  If every
    exponent $\eta_{i}$ is natural, then no derivative of lightlike
    geodesics diverges at $t_{0}$.  This is the case of course
    of a non-vanishing analytical $a(t)$ in the vicinity of $t_{0}$,
    such as, for instance, in de Sitter, $a(t)=\cosh t$, and anti-de
    Sitter universes, $a(t)=\cos t$.
    
    \item $\eta_{0}=-1/2$. For all of these cases $\ddot t$ does not
    blow up at $t_{0}$. Since $\dddot t\sim |t-t_{0}|^{\eta_{1}}$,
the third
    derivative  blows up if $\eta_{1}$ is negative.
    
    Since $t^{n)}\sim|t-t_{0}|^{\eta_{1}+(3-n)/2}$ in this case, if
    $\eta_{1}\in \big((n-4)/2,(n-3)/2\big)$, $n>2$, the derivative
$t^{n-1)}$
    does not blow up at $t_{0}$ but $t^{n)}$ does.
    
    \item $\eta_{0}=-2/3$.  In these models $\dddot t$ is finite at
    $t_{0}$.  Since $\ddddot t\sim |t-t_{0}|^{\eta_{1}+1/3}$, the
fourth derivative
    blows up if $\eta_{1}<-1/3$.
    
    Since $t^{n)}\sim|t-t_{0}|^{\eta_{1}+(5-n)/3}$, if
    $\eta_{1}\in \big((n-6)/3,(n-5)/3\big)$, $n>3$, the derivative
$t^{n-1)}$
    does not blow up at $t_{0}$ but $t^{n)}$ does.
\end{itemize}

Therefore we see different levels of singular behavior at $t=t_{0}$
along lightlike geodesics.  As $\eta_{0}$ decreases to get closer to
the value $-1$, the regularity of the geodesic improves in the sense
that one has to go to derivatives of higher order to find singular
behavior.  This limit model has been named superphantom
\cite{dabrowski} in the case of flat power-law cosmologies with
$w=-5/3$.  For $\eta_{0}$ greater or equal than $-1$ these geodesics
do not even reach the event at $t_{0}$.

As we have already pointed out,  the geodesic equation for $r$ 
does not add any further information, as we see in the equation of the
orbit
\[\left|\frac{dR}{dt}\right|=\frac{\left|\dot R\right|}{\dot t}=\frac
{1}{a(t)}\simeq
\frac{|t-t_{0}|^{-\eta_{0}}}{c_{0}}\;,\]which is
integrable close to $t_{0}$ if $\eta_{0}<1$. That is, it adds no new
information, since cases with $\eta_{0}\ge 1$ have already been shown
to be 
singular.

\begin{table}
\begin{tabular}{|c|c|c|c|c|c|}
    \hline
   $\eta_{0}$ & $\eta_{1}$ & $\dot t$ & $\ddot t$ & $\dddot t$ &
$t^{n)}$  \\
    \hline \hline
   $(0,\infty) $ & $(\eta_{0},\infty)$ & $\infty$ & $\infty$ &
$\infty$ & $\infty$  \\
   \hline
  $0$ & $(0,1)$ & finite & $\infty$ & $\infty$ & $\infty$  \\
  \cline{2-6}
  & $(1,2)$ & finite & finite & $\infty$ & $\infty$  \\
  \cline{2-6}
 & $(2,3)$ & finite & finite & finite & $\infty$  \\
\hline
  $(-1/2,0)$ & $(\eta_{0},\infty)$ & finite & $\infty$ & $\infty$ &
$\infty$  \\
  \hline
 $-1/2$ & $(-1/2,0)$ & finite & finite & $\infty$ & $\infty$  \\
 \cline{2-6}
 & $(0,1/2)$ & finite & finite & finite & $\infty$  \\
\hline
$(-2/3,-1/2)$ & $(\eta_{0},\infty)$ & finite & finite & $\infty$ &
$\infty$  \\
\hline
$-2/3$ & $(-2/3,-1/3)$ & finite & finite & finite & $\infty$  \\
    \hline
    $\big(\frac{1-n}{n},\frac{2-n}{n-1}\big)$ & $(\eta_{0},\infty)$ &
finite &
   finite & finite & $\infty$  \\
    \hline
    $(-\infty, -1]$  & $(\eta_{0},\infty)$ & / & / & / & / \\
    \hline
\end{tabular}
    \caption{Derivatives of lightlike geodesics at $t_{0}$. A slash
    indicates  the cases where $t_{0}$ is never reached.
    }
    \end{table}

\section{Strength of singularities along lightlike
geodesics\label{stronglight}}

As we have shown in the previous section, the ``strength'' of
singularities decreases qualitatively as the exponent $\eta_{0}$
decreases. It would be interesting to check these differences in
behavior with the usual definitions of strong singularities. 

The idea of a strong singularity was first introduced by Ellis and
Schmidt \cite{ellis}. A singularity is meant to be strong if 
tidal forces exert a severe disruption on finite objects falling into
it.  There have been several attempts to provide a rigorous
mathematical definition for this idea.

The finite volume is considered to be spanned by three orthogonal
Jacobi fields which form an orthonormal basis with the velocity of the
incomplete geodesic.  According to Tipler \cite{tipler}, the
singularity is strong if the volume tends to zero as the geodesic
approaches the value of proper time where it meets its end.
Kr\'olak's definition \cite{krolak} is less restrictive, since it just
requires the derivative of the volume with respect to proper time to
be negative.  This definition has been further refined in
\cite{rudnicki}.

However, these definitions are meant for focusing gravitational
forces, that is, $R_{ij}u^iu^j$ must be non-negative for timelike and
lightlike observers with velocity $u$.  Hence, these definitions leave
out the possibility of Big Rip singularities, but they can be extended
to these cases just reversing signs.  For instance, for negative
$R_{ij}u^iu^j$ Tipler's definition requires a volume tending to
infinity as the geodesic meets its end and Kr\'olak's definition
requires positive derivative of the volume close to the end of the
geodesic. 

Fortunately, one is not to construct the basis of Jacobi fields in
order to check these definitions.  There are necessary and sufficient
conditions due to Clarke and Kr\'olak \cite{clarke} related to
integrals of Riemann tensor components along the incomplete geodesic.
They are not affected by the inclusion of Big Rip singularities since
they just require components of the Riemann tensor to blow up along
causal geodesics.  In order to apply them we need the components of
the Riemann tensor, 
\[\frac{R^t_{rtr}}{f^2}=\frac{R^t_{\theta
t\theta}}{r^2}=\frac{R^t_{\phi t\phi}}{r^2\sin^2\theta}=aa'' \;,\] \[
R^r_{ttr}=R^\theta_{tt\theta}=R^{\phi}_{tt\phi}=\frac{a''}{a}\;,\]
\begin{eqnarray*}\frac{R^r_{\theta
r\theta}}{r^2}\!&=&\!\frac{R^r_{\phi
r\phi}}{r^2\sin^2\theta}=-\frac{R^\theta_{rr\theta}}{f^2}=
\frac{R^\theta_{\phi\theta\phi}}{r^2\sin^2\theta}=-\frac{R^\phi_{rr\phi}}{f^2}\\&=&
-\frac{R^\phi_{\theta\theta\phi}}{r^2}=a'^2+k\;,\end{eqnarray*}
omitting the ones that can be obtained by the symmetries of the
tensor, and Ricci tensor components, \[R_{tt}=-\frac{3a''}{a}\;,\
\frac{R_{rr}}{f^2}=\frac{R_{\theta\theta}}{r^2}=\frac{R_{\phi\phi}}{r^2\sin^2\theta}=
aa''+2(a'^2+k)\;.\]

For FLRW models conditions are simpler, since the Weyl tensor vanishes
and therefore only conditions related to the Ricci tensor are
relevant.  According to \cite{clarke}, a lightlike geodesic meets a
strong singularity, according to Tipler's definition, at proper time
$\tau_{0}$ if and only if the integral of the Ricci tensor
\begin{equation}\label{suftipler}
 \int_{0}^{\tau}d\tau'\int_{0}^{\tau'}d\tau''R_{ij}u^{i}u^j
\end{equation}
diverges as $\tau$ tends to $\tau_{0}$.

For Kr\'olak's definition the condition is less restrictive: a
lightlike geodesic meets a strong singularity 
at proper time $\tau_{0}$ if and only if the integral
\begin{equation}\label{sufkrolak}
 \int_{0}^{\tau}d\tau'R_{ij}u^{i}u^j
 \end{equation}
diverges as $\tau$ tends to $\tau_{0}$.

In our case, the velocity of the
geodesic is
\[u^t=\dot t=\frac{P}{a}\;,\qquad u^r=\dot r=\pm\frac{P}{fa^2}\;,\]
and
therefore the component of the Ricci tensor measured along the
geodesic is
\[
R_{ij}u^iu^j=2P^2\left(\frac{a'^2+k}{a^4}-\frac{a''}{a^3}\right)\;.\]

At first order in our generalised power expansion, we have two cases, 
depending on whether the curvature term is leading or not,
\[R_{ij}u^iu^j\simeq
\frac{2P^2\eta_{0}}{c_{0}^2|t-t_{0}|^{2(\eta_{0}+1)}}
+\frac{2kP^2}{c_{0}^4|t-{t_{0}}|^{4\eta_{0}}}
+\cdots\;,\]

\begin{itemize}
    
    \item First,  the case $\eta_{0}\le-1$, which is
    complete, since lightlike geodesics never reach $t=t_0$, as we
    have seen. 
    
    \item For $\eta_{0}\in (-1,1) $, $k\neq 0$ or $\eta_{0}>-1$, $k=
0$, using (\ref{tlight}), 
\begin{eqnarray*}R_{ij}u^iu^j&\simeq&
\frac{2P^2\eta_{0}}{c_{0}^2}|t-t_{0}|^{-2(\eta_{0}+1)}\\&\simeq&
\frac{2\eta_{0}}{(\eta_{0}+1)^2}\frac{1}{|\tau-\tau_{0}|
^{2}}\;,\end{eqnarray*}
produces a logarithmic divergence with Tipler's definition and an
inverse power divergence with Kr\'olak's for $0\neq \eta_{0}>-1$
and therefore we have a strong singularity in these cases.

\item There is a subcase left, $\eta_{0}=0$, for which
the approximation at first order leaves
\begin{eqnarray*}
R_{ij}u^iu^j&\simeq&2P^2\left(\frac{k}{c_{0}^4}-\frac{c_{1}\eta_{1}
    (\eta_{1}-1)}{c_{0}^3|t-t_{0}|^{2-\eta_{1}}}
\right)\\&\simeq &\frac{2kP^2}{c_{0}^4}-
\frac{2P^{\eta_{1}}c_{1}\eta_{1}(\eta_{1}-1)}{c_{0}^{\eta_{1}+1}|\tau-\tau_{0}|^{2-\eta_{1}}}
\;,\end{eqnarray*}
which provides no divergent integral with Tipler's definition, since
$\eta_{1}>0$, but provides one with Kr\'olak's one if
$\eta_{1}\in(0,1)$. This generalises the result of \cite{flrw}, since 
there it was shown that sudden singularities, a special case with
$\eta_{0}=0$, $\eta_{1}>1$, were not in fact singularities.

For $\eta_{1}=1$ we have to use still another term of the
expansion,
\begin{eqnarray*}
R_{ij}u^iu^j&\simeq&2P^2\left(\frac{k+c_1^2}{c_{0}^4}
-\frac{c_{2}\eta_{2}(\eta_{2}-1)}{c_{0}^3|t-t_{0}|^{2-\eta_{2}}}
\right)\\&\simeq &\frac{2(k+c_{1}^2)P^2}{c_{0}^4}-
\frac{2P^{\eta_{2}}c_{2}\eta_{2}(\eta_{2}-1)}{c_{0}^{\eta_{2}+1}|\tau-\tau_{0}|^{2-\eta_{2}}}
\;,\end{eqnarray*} which shows that this subcase does not produce a
divergent integral since $\eta_{2}>1$.

\item For $\eta_{0}=1$,
\begin{eqnarray*}
R_{ij}u^{i}u^j&\simeq& 2P^2
\frac{c_{0}^2+k}{c_{0}^4}|t-t_{0}|^{-4}\;,
\end{eqnarray*}
we see that both integrals are divergent, since the exponent is
smaller than $-2$, unless $k=-1$, $c_{0}=1$, values for which 
the Ricci tensor vanishes at first order, because at this order it is
Milne model.  We are to resort then to the
next term, $\eta_{1}>1$, 
\begin{eqnarray*}
R_{ij}u^{i}u^j&\simeq&- 
\frac{2P^2\eta_{1}(\eta_{1}-3)c_{1}}{|t-t_{0}|^{5-\eta_{1}}}\\&\simeq&-
\frac{2^{(\eta_{1}-3)/2}P^{(\eta_{1}-1)/2}\eta_{1}(\eta_{1}-3)c_{1}}{|\tau-\tau_{0}|^{(5-\eta_{1})/2}}\;,
\end{eqnarray*}
which produces no divergent integral for Tipler's definition but for
Kr\'olak's one the singularity is strong if $\eta_{1}<3$.

We need another term, $\eta_{2}>3$, to check the regularity of the
$\eta_{1}=3$ subcase with Kr\'olak's definition,
\begin{eqnarray*}
R_{ij}u^{i}u^j\simeq-
\frac{2P^2\eta_{2}(\eta_{2}-3)c_{2}}{|t-t_{0}|^{5-\eta_{2}}}\;,\end{eqnarray*}
and we find that is similar to the $\eta_{1}$ contribution. Hence it
does not diverge for $\eta_{2}>3$.

\item Finally, in the cases with $k\neq 0$ and $\eta_{0}\ge1$ the
leading term in the Ricci tensor
component is the curvature one, 
\begin{eqnarray*}
R_{ij}u^iu^j&\simeq&\frac{2kP^2}{a^4}\simeq
\frac{2kP^2}{c_{0}^4|t-{t_{0}}|^{4\eta_{0}}}
\\&\simeq&
\frac{2k(1+\eta_{0})^{-4/(1+\eta_{0})}P^{2-4\eta_{0}/(1+\eta_{0})}}{c_0^{4-4\eta_{0}/(1+\eta_{0})}
|\tau-{\tau_{0}}|^{4\eta_{0}/(1+\eta_{0})}},\end{eqnarray*}
which also provides divergent integrals since the exponent of the
denominator is larger than 2. Hence, the singularities are strong in
these cases either.

\end{itemize}

\begin{table}
\begin{tabular}{|c|c|c|c|c|c|}
    \hline
    ${\eta_{0}}$ & ${\eta_{1}}$ & ${k}$ & $c_{0}$ &\textbf{Tipler} &
    \textbf{Kr\'olak}  \\
    \hline\hline
    $(-\infty,-1]$ &  &   & & Regular & Regular  \\
    \cline{1-1} \cline{5-6}
	$(-1,0)$ & $(\eta_{0},\infty)$ & $0,\pm 1$ &$(0,\infty)$ & Strong &
Strong  \\
    \cline{1-2} \cline{5-6}
    $0$ & $(0,1)$ &  && Weak 
    & Strong  \\
	\cline{2-2} \cline{5-6}
	 & $[1,\infty)$ &   && Weak 
	 & Weak 
	  \\
	 \cline{1-2} \cline{5-6}
    $(0,1)$ & $(\eta_{0},\infty)$ &   && Strong & Strong  \\
    \cline{1-3} \cline{5-6}
    $1$ & $(1,\infty)$ & $0,1$  && Strong & Strong  \\
    \cline{2-6}
     & $(1,\infty)$ &  &$(0,1)\cup(1,\infty)$ & Strong & Strong  \\
     \cline{2-2} \cline{4-6}
     & $(1,3)$ & $-1$ & 1 &Weak
     & Strong  \\
     \cline{2-2} \cline{5-6}
     & $[3,\infty)$ &  &  &Weak 
     & Weak 
     \\
\hline
    $(1,\infty)$ & $(\eta_{0},\infty)$ & $0,\pm 1$  &$(0,\infty)$&
Strong & Strong  \\
    \hline

\end{tabular}
\caption{Degree of singularity of null geodesics around
$t_{0}$.
}
\end{table}

We conclude that for $\eta_{0}>-1$ lightlike geodesics in all models
meet a strong singularity except for the cases $\eta_{0}=0$ (and
$\eta_{1}\ge 1$ with Kr\'olak's definition) and $\eta_{0}=1$, $k=-1$,
$c_{0}=1$ ($\eta_{1}\ge 3$ with Kr\'olak's definition), which are
regular.  These are the only cases, together with $\eta_{0}\le -1$,
that are null geodesically complete, even though the curvature is
singular also for these models, as it was shown in \cite{visser}.

We notice that the different behavior of geodesics for positive and
negative $\eta_{0}$ does not quite influence the strength of the
curvature singularity at $t_{0}$.  Generically models with a Big Rip
have null geodesics with derivatives that do not blow up at $t_{0}$
whereas
all derivatives of null geodesics in models with a Big Bang or Crunch 
are infinite.

\section{Timelike geodesics\label{time}}

For timelike geodesics the relevant equation is, at first order of
the power expansion,
\begin{equation}\label{ttime}
\dot t=\sqrt{1+\frac{P^2}{a^2}}\simeq
\sqrt{1+\frac{P^2}{c_{0}^2}(t-t_{0})^{-2\eta_{0}}}
\;,\end{equation}
which can be solved explicitly in terms of hypergeometric functions,
\[(t-t_{0})F\left(\frac{1}{2},-\frac{1}{2p},1-\frac{1}{2p};
-\frac{P^2}{c_{0}^2}(t-t_{0})^{-2\eta_{0}}\right)\simeq\tau-\tau_{0}\;,\]
where $F$ is the hypergeometric function, but we shall not use this
expression.

It is clear, as it happened for lightlike geodesics, that for
$\eta_{0}>0$ the geodesic is singular at $t=t_{0}$, since $\dot t$
blows up there, unless $P$ is zero, which is a trivial regular
case,
\[t-t_{0}=\tau-\tau_{0}\;,\quad r=r_{0}\;,\]
the comoving congruence of fluid wordlines.

On the contrary, this derivative is well defined and
takes the value one if $\eta_{0}$ is negative. In this case we may
approximate $\dot t$ in the vicinity of $t_{0}$,
\begin{equation}
\dot t
\simeq 1+\frac{P^2}{2c_{0}^2}(t-t_{0})^{-2\eta_{0}}\;.\end{equation}

In order to carry out
the analysis of the behavior of these geodesics, we need expressions
for higher derivatives of coordinate time $t$,
\begin{subequations}
    \begin{eqnarray}
\ddot t&=&-\frac{P^2 a'}{a^3\sqrt{1+P^2 a^{-2}}}\dot
t=-\frac{P^2 a'}{a^3}\nonumber\\&\simeq& -\frac{P^2\eta_{0}
(t-t_{0})^{-2\eta_{0}-1}}{c_{0}^2}\;,\\ \dddot t&=&
P^2\left(\frac{3a'^2}{a^4}-\frac{a''}{a^3}\right)\dot t\\
&\simeq& \frac{P^2\eta_{0}(2\eta_{0}+1)
(t-t_{0})^{-2\eta_{0}-2}}{c_{0}^2}\;,\\
t^{n)}&\simeq&\tilde \lambda_{n}
(t-t_{0})^{-2\eta_{0}-n+1}\;.\nonumber
	\end{eqnarray}
with
\begin{equation}	 \tilde \lambda_{n}
=\frac{(-1)^{n+1}P^22\eta_{0}\cdots(2\eta_{0}+n-2)}{2c_{0}^2}\;.
\end{equation}

\end{subequations}

From these expressions we may draw valuable information about timelike
geodesics around $t_{0}$:

\begin{itemize}
    \item  $\eta_{0}>0$: As it happened in the lightlike case, 
    timelike geodesics are singular at $t_{0}$, since the derivative
    $\dot t$ blows up. For flat power-law models it corresponds to
$w>-1$.

    \item $\eta_{0}\in (-1/2,0)$: Again as it happened for lightlike
    geodesics, the derivative $\ddot t$ blows up at $t_{0}$ whereas
    $\dot t$ does not.  It corresponds to $w<-7/3$ for flat power-law
    models.

    \item  $\eta_{0}\in (-1,-1/2)$: For these cases we find the first
    difference with lightlike geodesics. The derivative
$\dddot t$ blows up at $t_{0}$, but
    $\ddot t$ does not. They correspond to flat power-law models with 
    $w\in (-7/3,-5/3)$ .
    
    \item $\eta_{0}\in \left(\frac{1-n}{2},\frac{2-n}{2}\right)$,
$n>1$: Whereas lightlike geodesics 
    did not reach $t_{0}$ in finite proper time in these models,
    timelike geodesics do, with regular $t^{n-1)}$ but with infinite
    $t^{n)}$ at $t_{0}$. The corresponding flat power-law
coefficients would 
    be $w\in \left(\frac{3n-2}{3(2-n)},\frac{3n+1}{3(1-n)}\right)$.
\end{itemize}

Again for the limit cases $\eta_{0}=\frac{1-n}{2}$ $(0,-1/2,-1\ldots)$
the contribution of the term with exponent $\eta_{0}$ to the
derivative $t^{n+1)}$ vanishes and we have to resort to the next term
in the expansion with non-vanishing contribution to higher
derivatives,
\begin{eqnarray}
\dot t&\simeq& 1+\frac{P^2}{2c_{0}^2}(t-t_{0})^{-2\eta_{0}}
-\frac{P^2c_{1}}{c_{0}^3}(t-t_{0})^{\eta_{1}-3\eta_{0}}
\nonumber\\&-&
\frac{P^4}{8c_{0}^4}(t-t_{0})^{-4\eta_{0}}+\cdots\;,
\end{eqnarray}
which is the term with exponent $\eta_{1}-3\eta_{0}$ and
therefore the relevant contribution to the derivative $t^{n+1)}$ is
of the form $(t-t_{0})^{\eta_{1}-3\eta_{0}-n}$.

Let us analyse some of these cases:

\begin{itemize}
    \item $\eta_{0}=0<\eta_{1}$: The discussion is entirely similar to
    the one for lightlike geodesics.  The scale factor does not vanish
    at $t_{0}$ and therefore these are sudden or weak singularities.
    Since $\dot t\sim |t-t_{0}|^{\eta_{1}}$, they have finite $\dot t$
    at $t_{0}$ and the derivative $\ddot t$ is finite for $\eta_{1}\ge
    1$.  If $\eta_{1}\in \big(n-2,n-1\big)$, $n>2$, the derivative
    $t^{n-1)}$ is also finite at $t_{0}$ but $t^{n)}$ is not.
    
    Again, if $\eta_{1}$ is natural, we would have to resort to the
    first exponent $\eta_{N}$ which is not natural and if all of them
   are natural, then no derivative of timelike geodesics diverges at
$t_{0}$. 
	
    \item $\eta_{0}=-1/2$.  In these cases $\ddot t$ is finite at
    $t_{0}$.  Since $\dddot t\sim |t-t_{0}|^{\eta_{1}-1/2}$, the third
    derivative is finite if $\eta_{1}\ge 1/2$.  When $\eta_{1}\in
    \big(n-7/2,n-5/2\big)$, $n>3$, the derivative $t^{n-1)}$ is
finite at
    $t_{0}$ and $t^{n)}$ is not.
   
   \item $\eta_{0}=-1$.  Now  $\dddot t$ is finite at
   $t_{0}$ and $\ddddot t\sim |t-t_{0}|^{\eta_{1}}$. Hence the fourth
derivative 
   is finite if $\eta_{1}$ is positive. For $\eta_{1}\in \big
(n-5,n-4\big)$,
   $n>4$, the derivative
    $t^{n-1)}$ is finite at $t_{0}$ whereas $t^{n)}$ is not.
\end{itemize}

\begin{table}
\begin{tabular}{|c|c|c|c|c|c|}
    \hline
   $\eta_{0}$ & $\eta_{1}$ & $\dot t$ & $\ddot t$ & $\dddot t$ &
$t^{n)}$  \\
    \hline \hline
   $(0,\infty) $ & $(\eta_{0},\infty)$ & $\infty$ & $\infty$ &
$\infty$ & $\infty$  \\
   \hline
  $0$ & $(0,1)$ & finite & $\infty$ & $\infty$ & $\infty$  \\
  \cline{2-6}
  & $(1,2)$ & finite & finite & $\infty$ & $\infty$  \\
  \cline{2-6}
 & $(2,3)$ & finite & finite & finite & $\infty$  \\
\hline
  $(-1/2,0)$ & $(\eta_{0},\infty)$ & finite & $\infty$ & $\infty$ &
$\infty$  \\
  \hline
 $-1/2$ & $(-1/2,1/2)$ & finite & finite & $\infty$ & $\infty$  \\
 \cline{2-6}
 & $(1/2,3/2)$ & finite & finite & finite & $\infty$  \\
\hline
$(-1,-1/2)$ & $(\eta_{0},\infty)$ & finite & finite & $\infty$ &
$\infty$  \\
\hline
$-1$ & $(-1,0)$ & finite & finite & finite & $\infty$  \\
    \hline
    $\big(\frac{1-n}{2},\frac{2-n}{2}\big)$ & $(\eta_{0},\infty)$ &
finite &
   finite & finite & $\infty$  \\
    \hline
\end{tabular}
    \caption{Derivatives of timelike geodesics at $t_{0}$}
    \end{table}

Summarizing, geodesic behavior is similar for timelike and lightlike
geodesics in models with $\eta_{0}>-1/2$, but there is a different
pattern for the rest of the models.  Differentiability of timelike
geodesics improves as $\eta_{0}$ decreases, but there are only
isolated cases for which they are completely regular and this makes a
difference with the lightlike case.  There are no timelike geodesics
which take an infinite proper time to reach $t_{0}$, as it happens with
null geodesics with $\eta_{0}\le -1$.

The equation (\ref{geods2}) for $r$ does not add further information
on
the behavior of timelike geodesics either. We may tackle the equation
of the orbit of 
the geodesics,
\[\left|\frac{dR}{dt}\right|=\frac{\left|\dot R\right|}{\dot t}=
\frac{P}{a(t)\sqrt{P^2+a^2(t)}}\;,\]in a similar fashion.

If $\eta_{0}>0$, we get, close to $t_{0}$,
\[\left|\frac{dR}{dt}\right|\le\frac{1}{ a(t)}\simeq
\frac{1}{ c_{0}}|t-t_{0}|^{-\eta_{0}}\;,\]
is not integrable if $\eta_{0}\ge 1$, but these are all already
singular 
cases.

If $\eta_{0}\le0$, we get, close to $t_{0}$,
\[\left|\frac{dR}{dt}\right|\le\frac{P}{ a^2(t)}\simeq
\frac{P}{ c_{0}^2}|t-t_{0}|^{-2\eta_{0}}\;,\]
which is integrable for $\eta_{0}<1/2$.

Therefore, no new singular behavior appears on considering the
geodesic equation for $r$. The radial coordinate is singular where $t$
is already singular.

\section{Strength of singularities along timelike geodesics
\label{strongtime}}

Again, it would be quite interesting to know whether the singularities
encountered by timelike geodesics are strong or not according to the
usual definitions.

Conditions like (\ref{suftipler}-\ref{sufkrolak}) are not so simple
for timelike geodesics, since there are no both necessary and
sufficient conditions in this case.  Those conditions become just
sufficient if the Weyl tensor vanishes.

A timelike geodesic meets a
strong singularity, according to Tipler's definition, at proper time
$\tau_{0}$ if the integral of the Ricci tensor
\begin{equation}\label{suftipler1}
 \int_{0}^{\tau}d\tau'\int_{0}^{\tau'}d\tau''R_{ij}u^{i}u^j
 \end{equation}
diverges as $\tau$ tends to $\tau_{0}$.

In contrast, for Kr\'olak's definition, a timelike geodesic meets a
strong singularity at proper time $\tau_{0}$ if the integral
\begin{equation}\label{sufkrolak1}
 \int_{0}^{\tau}d\tau'R_{ij}u^{i}u^j
 \end{equation}
diverges as $\tau$ tends to $\tau_{0}$.

First we use the comoving fluid worldline congruence, with velocity
$u=\partial_{t}$.  In this case, $\dot t=1$, proper time and
coordinate time are the same $t-t_{0}=\tau-\tau_{0}$.  The component
of the Ricci tensor measured by observers along this congruence,
\[R_{ij}u^i u^j=-\frac{3a''}{a}\simeq-
\frac{3\eta_{0}(\eta_{0}-1)}{|t-t_{0}|^{2}}=-
\frac{3\eta_{0}(\eta_{0}-1)}{|\tau-\tau_{0}|^{2}}\;,\] produces a
logarithmic divergence with Tipler's definition and an inverse power
divergence with Kr\'olak's one, so we may conclude that the
singularities are strong for all models with $1\neq \eta_{0}\neq 0$.
Therefore we have:

\begin{itemize}
    \item  For $1\neq \eta_{0} \neq 0$ the geodesics in the fluid
congruence 
    meet a strong singularity at $t_{0}$.

    \item  For $\eta_{0}=0$, we need another term in the expansion,
    \[\frac{3a''}{a}\simeq
\frac{3c_{1}\eta_{1}(\eta_{1}-1)}{c_{0}|t-t_{0}|^{2-\eta_{1}}}=
\frac{3c_{1}\eta_{1}(\eta_{1}-1)}{c_{0}|\tau-\tau_{0}|^{2-\eta_{1}}}\;,\]
and we see that in these cases the integrals do not diverge with
Tipler's definition, but they do with Kr\'olak's for
$\eta_{1}\in(0,1)$.
    \item  For $\eta_{0}=0$, $\eta_{1}=1$, we still need another term,
    \[\frac{3a''}{a}\simeq
\frac{3c_{2}\eta_{2}(\eta_{2}-1)}{c_{0}|t-t_{0}|^{2-\eta_{2}}}=
\frac{3c_{2}\eta_{2}(\eta_{2}-1)}{c_{0}|\tau-\tau_{0}|^{2-\eta_{2}}}\;,\]
in order to check that integrals do not diverge for these models with 
both definitions, since $\eta_{2}>1$.

\item For $\eta_{0}=1$, we resort to the second term in the expansion,
\[\frac{3a''}{a}\simeq
\frac{3c_{1}\eta_{1}(\eta_{1}-1)}{c_{0}|t-t_{0}|^{3-\eta_{1}}}=
\frac{3c_{1}\eta_{1}(\eta_{1}-1)}{c_{0}|\tau-\tau_{0}|^{3-\eta_{1}}}\;,\]
and we find that for these models the integrals do not diverge with
Tipler's definition, but they do with Kr\'olak's for
$\eta_{1}\in(1,2]$.
\end{itemize}

\begin{table}
\begin{tabular}{|c|c|c|c|}
    \hline
    ${\eta_{0}}$ & ${\eta_{1}}$  & \textbf{Tipler} &
    \textbf{Kr\'olak}  \\
    \hline\hline
    $(-\infty,0)$ & $(\eta_{0},\infty)$ & Strong & Strong  \\
	\hline
	$0$ & $(0,1)$  & Weak & Strong  \\
	\cline{2-4}
	 & $[1,\infty)$  & Weak & Complete  \\
    \hline
    $(0,1)$ & $(\eta_{0},\infty)$ & Strong & Strong  \\
    \hline
    $1$ & $(1,2]$ &  Weak & Strong  \\
    \cline{2-4}
     & $(2,\infty)$ &  Weak & Weak  \\
\hline
    $(1,\infty)$ & $(\eta_{0},\infty)$ &  Strong & Strong  \\
    \hline
\end{tabular}
\caption{
 Degree of singularity of  the fluid congruence  of timelike
geodesics around $t_{0}$.}
\end{table}

This result may be further refined using timelike radial geodesics,
for which
\[u^t=\dot t=\sqrt{1+\frac{P^2}{a^2}}\;,\qquad u^r=\dot
r=\pm\frac{P}{fa^2}\;,\] 
\[R_{ij}u^i u^j=-\frac{3a''}{a}+2P^2\left(\frac{a'^2+k}{a^4}-\frac{a''}{a^3}
    \right)\;.\]
    
Taking a look at the geodesic equation (\ref{ttime}) for $t$, we
notice three different possibilities:

\begin{itemize}
    \item  $\eta_{0}<0$: Since $\dot t\simeq 1$,
$t-t_{0}\simeq\tau-\tau_{0}$
    close to $t_{0}$.

    \item  $\eta_{0}=0$: Now $\dot t\simeq
\sqrt{1+P^2/c_{0}^2}=\alpha$ close to
$t_{0}$ and so $t-t_{0}\simeq\alpha(\tau-\tau_{0})$.

    \item  $\eta_{0}>0$: For these cases $\dot t\simeq P/a$ close to
$t_{0}$ as for lightlike geodesics.
\end{itemize}

Accordingly, there are several cases:

\begin{itemize}
\item $\eta_{0}<0$: At lowest order, the $P\!\!-{\rm dependent}$ terms
\[\frac{a'^2+k}{a^4}-\frac{a''}{a^3}\simeq
\frac{\eta_{0}}{c_{0}^2|\tau-\tau_{0}|^{2(\eta_{0}+1)}}\;,\]
produce no divergent integral with Tipler's definition but they
do with Kr\'olak's one for $\eta_{0}\in [-1/2,0)$, but it does not
matter, since the first term was already seen to be divergent, as it
is
the same as for the fluid congruence in all these cases.

\item $\eta_{0}=0$: The $P\!\!-{\rm dependent}$ term is essentially 
the same as for lightlike geodesics and we reach therefore the 
same conclusion: these models produce no divergent integral 
with Tipler's definition, but with Kr\'olak's one they do if
$\eta_{1}\in(0,1)$.
The same happens with the first term, which is the same as for the
fluid congruence.

\item $\eta_{0}=1$: The $P\!\!-{\rm dependent}$ term is the same as for
lightlike
geodesics.  Hence, these cases are all singular but except maybe for
$k=-1$, $c_{0}=1$.  Models with $k=-1$, $c_{0}=1$ and $\eta_{1}<3$
are singular with
Kr\'olak's definition. On the other hand, the first term,
\[\frac{3a''}{a}\simeq
\frac{3c_{1}\eta_{1}(\eta_{1}-1)}{c_{0}|t-t_{0}|^{3-\eta_{1}}}\simeq
\left(\frac{2P}{c_{0}}\right)^{\frac{\eta_{1}-3}{2}}\frac{3c_{1}\eta_{1}(\eta_{1}-1)}{c_{0}
|\tau-\tau_{0}|^{\frac{3-\eta_{1}}{2}}},\] does not diverge.

\item $1\neq \eta_{0}>0$: The $P\!\!-{\rm dependent}$ term for these
geodesics is the same
as for lightlike geodesics and therefore it is divergent in all cases,
though the first term,
\[\frac{3a''}{a}\simeq
\frac{3\eta_{0}(\eta_{0}-1)}{|t-t_{0}|^{2}}\simeq
\left(\frac{c_{0}}{P(1+\eta_{0})}\right)^{\frac{2}{1+\eta_{0}}}
\frac{3\eta_{0}(\eta_{0}-1)}{|\tau-\tau_{0}|^{\frac{2}{1+\eta_{0}}}},\]

does not diverge with Tipler's definition and only with Kr\'olak's for
$\eta_{0}\in(0,1)$.

\end{itemize}

Therefore we have so far exactly the same models with strong
singularities as we found for lightlike geodesics plus the
$\eta_{0}\le -1$ models, which are null, but not timelike,
geodesically complete.  That is, we know that all models with $0\neq
\eta_{0}\neq 1$ have strong singularities at $t_{0}$.  We don't know
what happens with models with $\eta_{0}=0$, though those with
$\eta_{1}\in (0,1)$ have strong singularities according to Kr\'olak.
And the same happens with models with $\eta_{0}=1$, $k=-1$, $c_{0}=1$,
though those with $\eta_{1}\in (1,3)$ have also strong singularities
according to Kr\'olak.

\begin{table}
    \begin{tabular}{|c|c|c|c|c|c|}
	\hline
	${\eta_{0}}$ & ${\eta_{1}}$ & ${k}$ & $c_{0}$ &\textbf{Tipler} &
	\textbf{Kr\'olak}  \\
	\hline\hline
	$(-\infty,0)$ & $(\eta_{0},\infty)$ &   && 
	Strong & Strong  \\
	\cline{1-2} \cline{5-6}
	$0$ & $(0,1)$ &  && Weak & Strong  \\
	    \cline{2-2} \cline{5-6}
	     & $[1,\infty)$ & $0,\pm 1$  &$(0,\infty)$& Weak & Weak
\\
	     \cline{1-2} \cline{5-6}
	$(0,1)$ & $(\eta_{0},\infty)$ &   && Strong & Strong  \\
	\cline{1-3} \cline{5-6}
	$1$ & $(1,\infty)$ & $0,1$  && Strong & Strong  \\
	\cline{2-6}
	 & $(1,\infty)$ &  &$(0,1)\cup(1,\infty)$ & Strong & Strong  \\
	 \cline{2-2} \cline{4-6}
	 & $(1,3)$ & $-1$ & 1 & Weak & Strong  \\
	 \cline{2-2} \cline{5-6}
	 & $[3,\infty)$ &  &  &Weak & Weak  \\
    \hline
	$(1,\infty)$ & $(\eta_{0},\infty)$ & $0,\pm 1$  &$(0,\infty)$&
Strong & Strong \\
	\hline
    \end{tabular}
\caption{ Degree of singularity of radial timelike geodesics
around $t_{0}$}
\end{table}

Since  the condition on
integrals of the Ricci tensor is not also a necessary condition for
the
appearance of strong singularities, we have to check other ways to get
information about the $\eta_{0}=0$ and $\eta_{0}=1$ models.

For Tipler's definition \cite{clarke}, if a causal geodesic with
velocity $u$ meets a strong
singularity, then the integral
\begin{equation} I^i_{j}(\tau)=\int_{0}^\tau d\tau'\int_{0}^{\tau'}
    d\tau''\left|R^i_{kjl}u^ku^l\right|\;,\end{equation}
diverges as $\tau$ tends to $\tau_{0}$ for some $i$, $j$. The
components are referred to a parallely transported orthonormal frame.

Again Kr\'olak's definition is less restrictive and just requires
that the integral
\begin{equation} I^i_{j}(\tau)=\int_{0}^\tau d\tau'
    \left|R^i_{kjl}u^ku^l\right|\;,\end{equation}
diverges as $\tau$ tends to $\tau_{0}$ for some $i$, $j$.

We begin again with the fluid worldline congruence, $u^t=1$, for which
the only non-vanishing components of the Riemann tensor,
\[R^i_{tit}=-\frac{a''}{a}\;,\quad i=r,\theta,\phi\;,\] produce a
necessary condition which is the same as the already studied
sufficient condition. 

Therefore, geodesics in the fluid worldline congruence meet a strong
singularity if and only if $1\neq \eta_{0}\neq 0$ (or $\eta_{0}=0$,
$\eta_{1}\in (0,1)$ and $\eta_{0}=1$, $\eta_{1}\in (1,2)$ with
Kr\'olak's
definition).

Radial timelike geodesics show more strong singularities, as we may
see.  We complete the orthonormal basis formed by $u$, \[u=\dot
t\partial_{t}+\dot r\partial_{r}=
\sqrt{1+\frac{P^2}{a^2}}\partial_{t}\pm\frac{P}{a^2f}\partial_{r}\;,\]

and a vector $v$, \[v=af\dot r\,\partial_{t}+\frac{\dot
t}{af}\partial_{r}=\pm\frac{P}{a}\partial_{t}+\frac{1}{af}\sqrt{1+\frac{P^2}{a^2}}\partial_{r}\;,\]
adding the corresponding unitary vectors parallel to
$\partial_{\theta}$ and $\partial_{\phi}$.  The parallel transport
requirement is trivially satisfied.

The $\theta$ and $\phi$ components of the Riemann tensor,
\begin{eqnarray*}
R^{\theta}_{k\theta l}u^ku^l&=&R^{\theta}_{t\theta t}\dot t^2+
R^{\theta}_{r\theta r}\dot r^2+\left(R^{\theta}_{t\theta r}+
R^{\theta}_{r\theta t}\right)\dot t\dot
r\\&=&-\frac{a''}{a}-P^2\left\{
\frac{a''}{a^3}-\frac{a'^2+k}{a^4}\right\}=R^{\phi}_{k\phi
l}u^ku^l\;,\end{eqnarray*} have similar terms as $R_{ij}u^i u^j$ and
therefore produce the same results as the corresponding sufficient
condition.

Finally, the $v$ components,
\begin{eqnarray*}R_{ikjl}v^iu^kv^ju^l&=&R_{trtr}\left\{
\frac{P^4}{a^6f^2}+\frac{1}{a^2f^2}\left(1+\frac{P^2}{a^2}\right)^2\right..\\&-&
\left.2\frac{P^2}{a^4f^2}\left(1+\frac{P^2}{a^2}\right)\right\}=-\frac{a''}{a}\;,\end{eqnarray*}
provide a term that has already been discussed and therefore we may
conclude also that sufficient conditions for the appearance of strong 
singularities along timelike geodesics are also necessary, as it
happened for lightlike ones.

\section{Discussion \label{discuss}}

We have obtained a thorough classification of singular events in FLRW
cosmological models in terms of the exponents of a generalised power
expansion of the scale factor in coordinate time around a cosmological
milestone at $t_{0}$.  The behavior of causal geodesics has been
obtained in the vicinity of the event.  The first difference that has
been found is that whereas the velocity of causal geodesics blows up
at Big Bang and Big Crunch singularities, it is finite at Big Rip
singularities, as well as acceleration and other derivatives,
depending on the first exponent in the expansion, $\eta_{0}$.  For sudden singularities the
velocity is finite and the acceleration may be finite or not, depending
on the next exponent, $\eta_{1}$.

However this difference of regularity between Big Bang/Crunch and Big
Rip singularities does not prevent the strong character of both types
of cosmological milestones with both Tipler and Kr\'olak's definitions
of strong singularities.  There is only a curious feature in Big Rip
singularities in models with $\eta_{0}\le -1$ (which are precisely
those favoured by observations): lightlike geodesics do not reach the
curvature singularity at $t_{0}$ in finite proper time and therefore
these spacetimes are null geodesically complete close to the singular
event.  Hence photons never experience Big Rip singularities and the
universe would last eternally for them.  This feature, however, is
lost on dealing with timelike geodesics, which reach $t_{0}$ in finite
proper time and meet a strong singularity.

The only models which allow regular behavior close to $t_{0}$ are
those with $\eta_{0}=0$ and with $\eta_{0}=1$, $k=-1$, $c_{0}=1$.  The
latter ones are Milne universes at first order, which are essentially
Minkowski spacetime after extending the model beyond $t_{0}$.  The
former ones include models with a non-vanishing analytical scale
factor, such as de Sitter universes, and models with sudden
singularities, which have finite velocity, but non-finite acceleration
or higher derivatives of the parametrization of the geodesics
depending on the exponent $\eta_{1}$.  The larger this exponent is,
the better the properties of the model are.  These cosmologies prevent
the formation of strong singularities according to Tipler's
definition, which requires the crushing to zero or disrupting to
infinity of finite volume objects evolving along causal geodesics.
With Kr\'olak's definition, which requires just a positive derivative
of volume for Big Bang and Big Rip singularities and a negative
derivative for Big Rip singularities, strong singularities are avoided
in models with $\eta_{1}\ge 1$.  This definition seems more
appropriate, since causal geodesics in models with $\eta_{1}\in (0,1)$
do not have finite acceleration and therefore geodesic equations would
be singular at $t_{0}$, though the curves may be extend beyond that
event.

We may compare these results with those studied by Catto\"en and Visser in \cite{visser}, where
just singularities in curvature were considered, without taking into
account their strength nor the behavior of causal geodesics.
 Those authors found that the only models without polynomial
curvature singularities are those with $\eta_{0}=0$, $\eta_{1}\ge 2$
or $\eta_{1}=1$, $\eta_{2}\ge 2$, and those with $\eta_{0}=1$, $k=-1$,
$c_{0}=1$, $\eta_{1}\ge 3$.  Dealing with derivative curvature
singularities, the list reduces to models with $\eta_{0}=0$ and
natural exponents $\eta_{i}$, $i\ge 1$ and those with $\eta_{0}=1$,
$k=-1$, $c_{0}=1$ and natural exponents $\eta_i\ge 3$, $i\ge 1$.
Derivative curvature singularities are not reflected in our
classification since derivatives of the Riemann tensor appear neither
in geodesic equations nor in Jacobi equations.  The apparent
discrepancy between our results and the presence of polynomial
curvature singularities lies on the fact that either geodesics do not
reach that singularity or that they reach it, but the curvature growth
is not enough to form a strong singularity.

Finally, another consequence is that singularities appear just in
models with vanishing, divergent or non-smooth scale factors.  From
the mathematical point of view at least it is worth mentioning that
regular models are an open set within the family of smooth homogeneous
and isotropic spacetimes, as it happened for instance with
inhomogeneous scalar field Abelian diagonal $G_{2}$ models
\cite{wide}.  On the contrary, singular models are not an open set,
since the vanishing requirement is not generic.

\section*{Acknowledgments}L.F.-J. is supported by the Spanish Ministry
of Education and Science Project FIS-2005-05198. R.L. is supported by
the University
of the Basque Country through research grant UPV00172.310-14456/2002
and by the Spanish Ministry of Education and Culture through the RyC
program, 
and research
grants FIS2004-01626 and FIS2005-
01181.

\end{document}